\documentclass[runningheads]{llncs}
\usepackage[T1]{fontenc}
\usepackage{amsmath}
\usepackage{algorithm}
\usepackage{epsfig}
\usepackage{subfigure}
\usepackage{graphics}
\usepackage{color,ulem}
\usepackage{amssymb}
\usepackage{algorithm,hyperref}
\usepackage[noend]{algpseudocode}
\usepackage[numbers,sort]{natbib}
\usepackage{comment}
\usepackage{graphicx}
\newcommand{\angstrom}{\textup{\AA}}

\begin{document}
\title{Machine Learning for First Principles Calculations of Material Properties for Ferromagnetic Materials \thanks{This manuscript has been authored in part by UT-Battelle, LLC, under contract DE-AC05-00OR22725 with the US Department of Energy (DOE). The US government retains and the publisher, by accepting the article for publication, acknowledges that the US government retains a nonexclusive, paid-up, irrevocable, worldwide license to publish or reproduce the published form of this manuscript, or allow others to do so, for US government purposes. DOE will provide public access to these results of federally sponsored research in accordance with the DOE Public Access Plan (\url{http://energy.gov/downloads/doe-public-access-plan}).}}
%
%
\author{Markus Eisenbach\inst{1}\orcidID{0000-0001-8805-8327} \and
Mariia Karabin\inst{1}\orcidID{0000-0003-0081-8497} \and
Massimiliano Lupo Pasini\inst{2}\orcidID{0000-0002-4980-6924} \and
Junqi Yin\inst{1}\orcidID{0000-0003-3843-5520}}
\authorrunning{M. Eisenbach et al.}
%
\institute{National Center for Computational Sciences, Oak Ridge National Laboratory, Oak Ridge TN 37831, USA 
\email{\{eisenbachm,karabinm,yinj\}@ornl.gov}\\
\and
Computational Sciences and Engineering Division, Oak Ridge National Laboratory, Oak Ridge TN 37831, USA 
\email{lupopasinim@ornl.gov}}
\maketitle              
\begin{abstract}
The investigation of finite temperature properties using Monte-Carlo (MC) methods requires a large number of evaluations of the system’s Hamiltonian to sample the phase space needed to obtain physical observables as function of temperature.
DFT calculations can provide accurate evaluations of the energies, but they are too computationally expensive for routine simulations. To circumvent this problem, machine-learning (ML) based surrogate models have been developed and implemented on high-performance computing (HPC) architectures. 
In this paper, we describe two ML methods (linear mixing model and HydraGNN) as surrogates for first principles density functional theory (DFT) calculations with classical MC simulations. 
These two surrogate models are used to learn the dependence of target physical properties from complex compositions and interactions of their constituents. 
We present the predictive performance of these two surrogate models with respect to their complexity while avoiding the danger of overfitting the model. An important aspect of our approach is the periodic retraining with newly generated first principles data based on the progressive exploration of the system’s phase space by the MC simulation.
The numerical results show that HydraGNN model attains superior predictive performance compared to the linear mixing model for magnetic alloy materials.

\keywords{Machine Learning  \and Surrogate Models \and Material Science, Solid Solution Alloys.}
\end{abstract}
\section{Introduction}
 The investigation of finite temperature properties using Monte-Carlo (MC) methods requires a large number of evaluations of the system’s Hamiltonian to sample the phase space needed to obtain physical observables as function of temperature. Density functional calculations can provide accurate evaluations of the energies, but they are too computationally expensive for routine simulations.
 Surrogate models can alleviate the computational cost to complete classical MC simulations \cite{Tetot1996, doi:10.1080/14786435.2013.805275, MOHAMMADI20209620, Walle2002SelfdrivenLM, Lavrentiev} since they can estimate finite temperature properties orders of magnitude faster than the original density functional theory (DFT) code.
 Although surrogate models generally do not attain the same accuracy as the physics-based model on which they are trained, a hybrid approach that combines initial inexpensive surrogate model evaluations with limited expensive DFT refinements can drastically reduce the computational time without significantly affecting the accuracy of the final inference with respect to running the DFT code throughout the entire classical MC workflow \cite{Reitz}. 
 
 The identification of an appropriate surrogate model must take into consideration multiple factors such as (i) the dimensionality of the data space that needs to be explored, (ii) the amount of training data available, (iii) the complexity of the cause/effect relation that connects input features and target properties, (iv) the available computational resources. 
 In situations similar to the ones addressed in this work where the training data is generated by running DFT calculations, (ii) and (iv) are strongly connected, as the computational resources available determine the amount of DFT data that can be generated for training. 
 
 Deep learning (DL) models have attracted a lot of attention in the material science community due to their ability to capture highly non-linear relations between input features and target properties \cite{CONDUIT2019107644, cgcnn, gatgnn, grain_boundary, 10.3389/fmats.2022.865270, lupo_gcnn, Lupo_Pasini_2022}. However, their success relies on the a subtle connection between the complexity of the DL model and the amount of available training data. 
 
In this paper we will investigate the capability of two different classes of surrogate models to describe the energy of a magnetic alloy system with sufficient accuracy needed to perform statistical mechanics investigations of alloy ordering.

We will first describe the two models we will be investigating, followed by a description of the physical alloy system, body-centered tetragonal FePt, that will provide the basis for our comparison. 
This will be followed by the results of our numerical experiments and finally we will discuss the results and conclude.
 
\section{Surrogate models}

\subsection{Linear model}
To model the energy of a refractory high entropy alloy system, the effective pair interactions (EPI) \cite{xianglin21} was proposed. The energy is approximated by the summation of local energies as follows, 
\begin{align}
    {\color{black}E \approx N \sum_{p'<p,m} V_m^{pp'} \Pi_m^{pp'} + V^p_i + V^0 + \epsilon,}
\label{eq:linear}
\end{align}
where $N$ is the total number of atoms, $V^0$ is the bias term same for all sites, $V_i^p$ is a single-site term depending only on the chemical component $p$ of atom $i$, and $\Pi_{m}^{pp'}$ is the proportion of $pp'$ interaction in the $m$-th neighboring shell. The EPI model has been shown \cite{xianglin21} to work well for non-magnetic alloy systems.

\subsection{Graph convolutional neural network}
\begin{figure}[htb!]
\centering
\includegraphics[width=\textwidth]{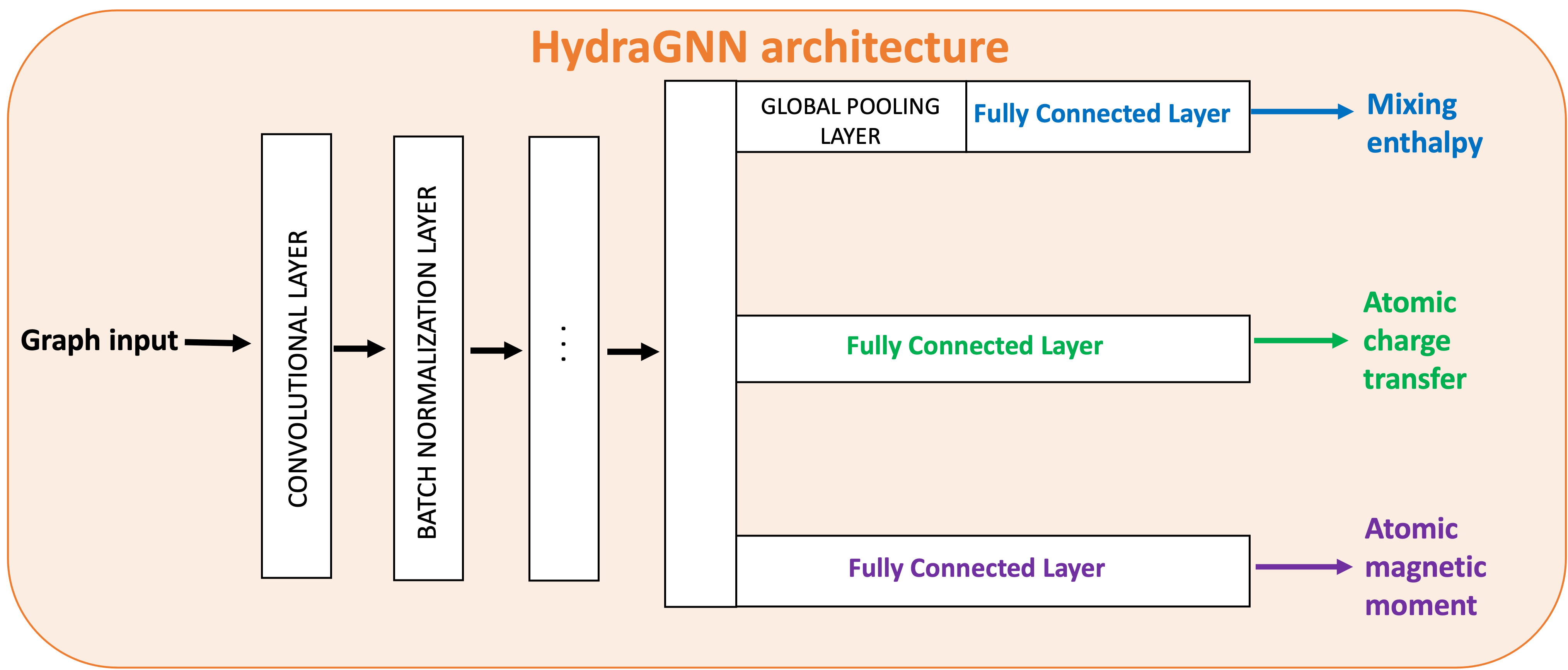}\\
\caption{HydraGNN architecture when used as a surrogate model for DFT calculations of mixing enthalpy.}
\label{fig:HydraGNN-architecture}
\end{figure}

GCNN models map atomic structure by naturally converting them into a graph, where atoms are interpreted as nodes and interatomic bonds are interpreted as edges, and outputs total (graph-level) and atomic (node-level) physical properties. 
The typical GCNN architecture is characterized by three different types of hidden layers: graph convolutional layers, graph pooling layers, and fully connected layers. 
The graph convolutional layers extract information from the graph samples about local features that model the short-range interactions of an atom with it neighbors. GCNNs embed the interactions between nodes (atoms) by representing the local interaction zone as a hyperparameter that cuts-off the interaction of a node with all the other nodes outside a prescribed local neighborhood. This is identical to the approximation made by many atomic simulation methods, including the LSMS-3 code used to generate the DFT training data, which ignore interactions outside a given cutoff range. Larger sizes of the local neighborhood lead to a higher computational cost to train the model, as the number of regression coefficients to train at each hidden convolutional layer increases 
proportional to the number of neighbors.
The second set of layers extracts global features that describe long range weak interactions between atoms far from each other in the crustal structure. The architecture of a GCNN model is shown schematically in Figure~\ref{fig:HydraGNN-architecture}. 

Our implementation of GCNN, called HydraGNN, uses \texttt{Pytorch} \cite{pytorch2019, pytorch} as both a robust NN library, as well as a performance portability layer for running on multiple hardware architectures. This enables HydraGNN to run on CPUs and GPUs, from laptops to supercomputers, including ORNL's Summit and NERSC's Perlmutter. The \texttt{Pytorch Geometric} \cite{fey_2019, torch_geometric} library built on \texttt{Pytorch} is particularly important for our work and enables many GCNN models to be used interchangeably. HydraGNN is openly available on GitHub \cite{hydragnn}.

A variety of graph convolutional layers, e.g., principal neighborhood aggregation (PNA) \cite{corso_principal_2020}, crystal GCNN (CGCNN) \cite{cgcnn} and GraphSAGE \cite{hamilton2017inductive}, have been developed.
Previous studies using GCNN models for solid solution alloys \cite{lupo_gcnn, Lupo_Pasini_2022} showed that PNA better discriminates different atomic configurations which in turn improves the final accuracy of the model. 
Therefore,  the Pytorch Geometric implementation of the PNA is used in this work as graph convolutional layer in HydraGNN.
Batch normalizations are performed between consecutive graph convolutional layers along with a ReLU activation function.
Graph pooling layers are connected to the end of the convolution-batch normalization stack to gather feature information from the entire graph to collapse the node feature into a single feature. This is achieved by summing the local interactions of each atom with its neighbors and use the result to estimate global properties such as the mixing enthalpy of an alloy.  
Fully connected (FC) layers are positioned at the end of the architecture to take the results of pooling, i.e. extracted features, and provide the output prediction. 

Further details on the behavior of HydraGNN with different sizes of the local neighborhood have been previously reported \cite{lupo_gcnn}.

\section{Ferromagnetic materials}

Itinerant ferromagnetic materials are typically metals. They exhibit magnetization even in the absence of a magnetic field. FePt is one of the examples of such a ferromagnetic material. The magnetic properties of this material are highly dependent on the chemical ordering and stoichiometry \cite{Vlaic2010}, according to which the magnetic state can change from ferromagnetic to antiferromagnetic\cite{Bacon1963}, a non-collinear spin structure \cite{Higashiyama2003}, or a mixture of antiferromagnetic orderings \cite{Tobita2010}. 
The total magnetic moment of FePt is presented in Figure~\ref{fig:correlation_plots}.
While pure Fe is magnetic and Pt is non-magnetic, the formation of magnetic moments in alloys is driven by the collective behavior of electrons in the alloy. Indeed there is a long history of trying to understand and describe the magnetism in materials dating back to the early days of quantum theory \cite{Slater1936}. The DFT calculations that form the basis of the present work take these collective electron behavior into account. In particular in the FePt systems, the magnetic moment associated with the Pt sites depend on their environment and the Fe concentration. This can be clearly seen in the center picture of Figure \ref{enthalpy} where the rate of change in the magnetic moment at Fe concentrations below $\approx 15\%$ depends on the amount of Fe in the system as the induced moments on the Pt site increases, whereas above this threshold the moments on the Pt sites have reached their saturated values and the total magnetization follows the Fe and Pt concentrations.

\subsection{Solid solution binary alloy dataset}
\label{dataset_section}

\begin{figure}[h]
    \centering
    \includegraphics[width=0.5\textwidth]{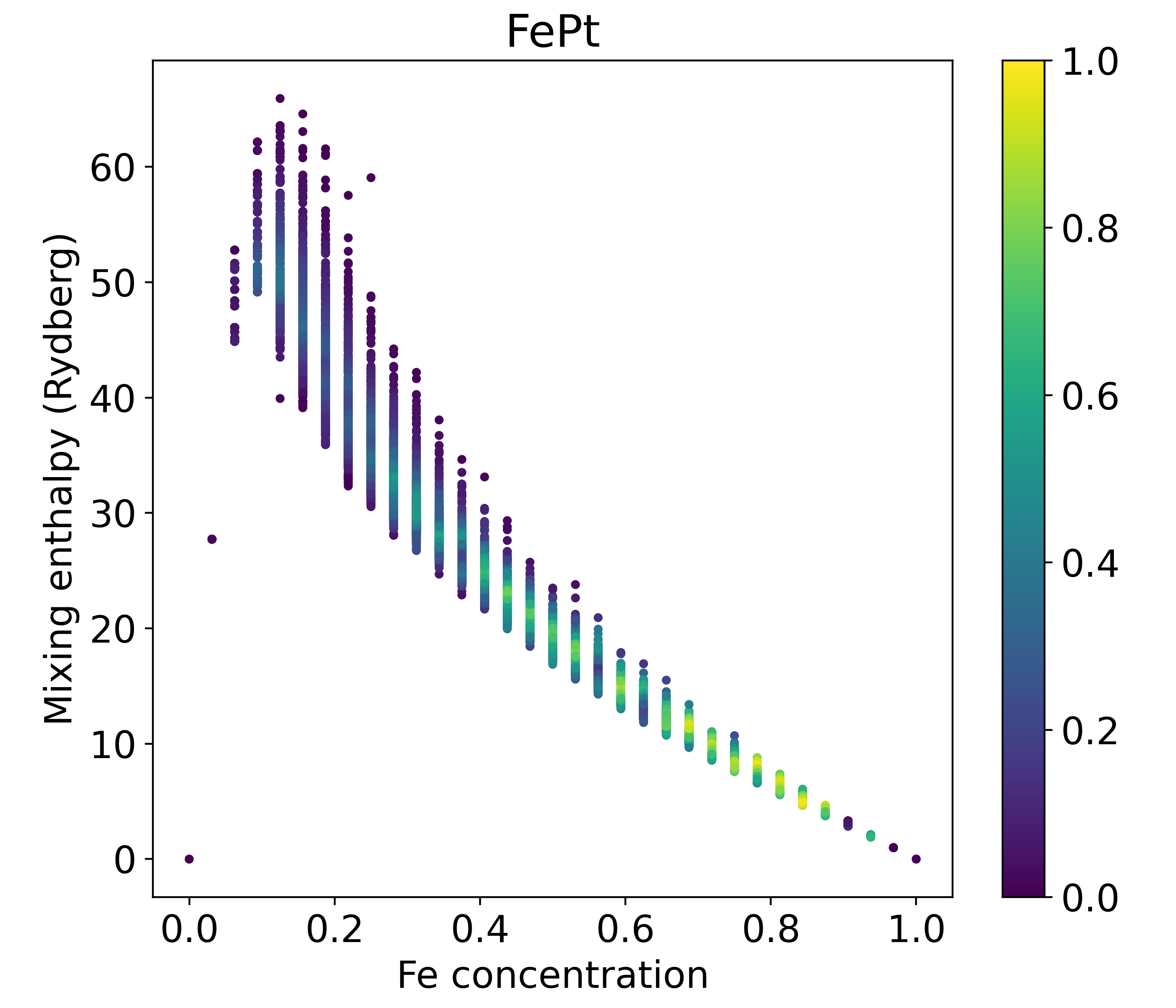}
    \includegraphics[width=0.5\textwidth]{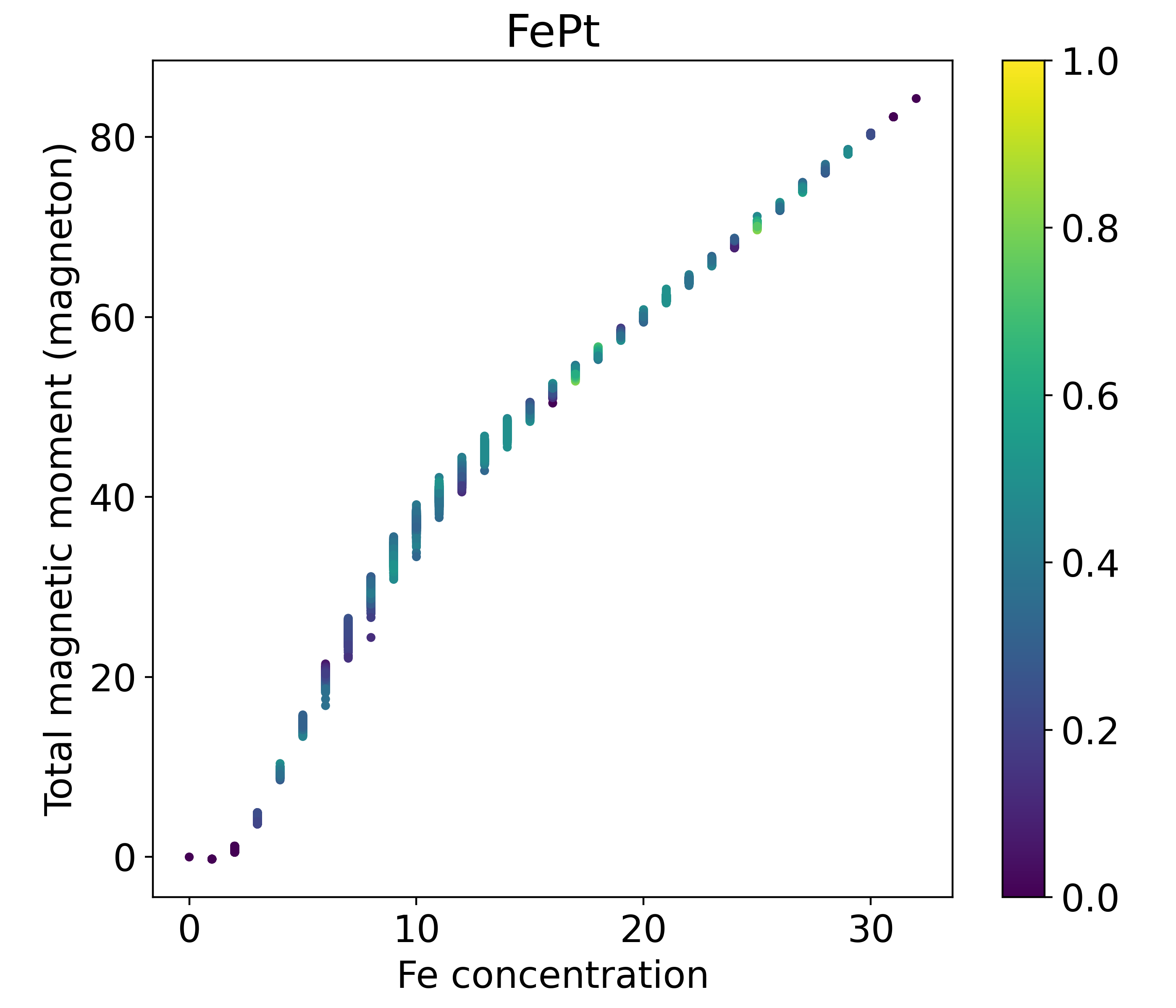}
    \includegraphics[width=0.5\textwidth]{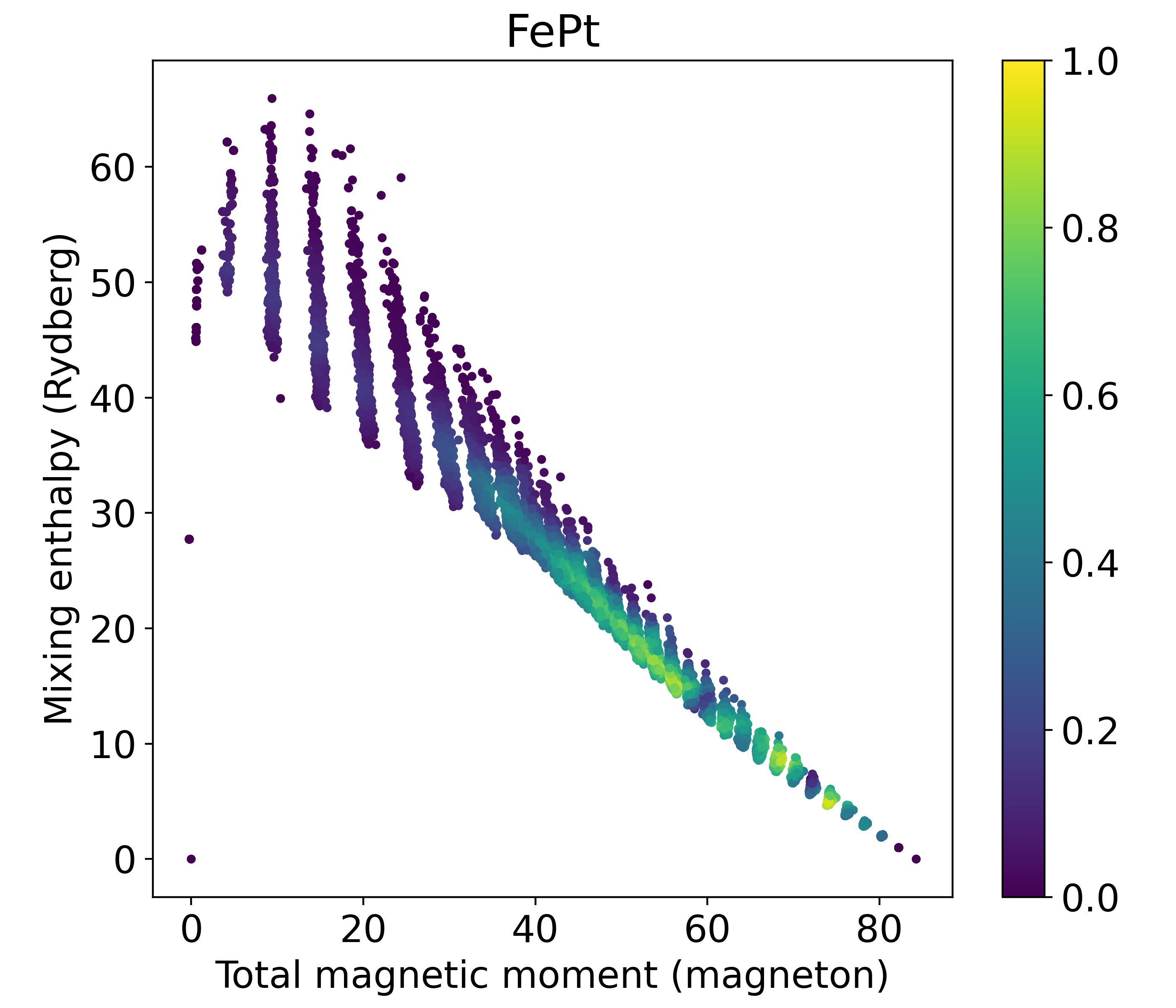}
    \caption{Top: configurational mixing enthalpy of solid solution binary alloy FePt with BCT structure as a function of Fe concentration. Center: scatter plot of total magnetic moment against concentration of iron. Bottom: scatter plot of mixing enthalpy against total magnetic moment (right). The color map in each plot indicates the relative frequency of data. }
    \label{enthalpy}
\end{figure}

In this work we focus on a solid solution binary alloy, where two constituent elements are randomly placed on an underlying crystal lattice. We use a dataset for FePt alloys available through the OLCF Constellation~\cite{FePt} which includes the total enthalpy, atomic charge transfer, and atomic magnetic moment. Each atomic sample has a body centered tetragonal (BCT) structure with a $2 \times 2 \times4$ supercell. The dataset was computed with LSMS-3 \cite{lsms-code}, a locally self-consistent multiple scattering (LSMS) DFT application \cite{Wang1995, Eisenbach2017}. The dataset was created with fixed volume in order to isolate the effects of graph interactions and graph positions for models such as GCNN. This produces non-equilibrium alloy samples, with non-zero pressure and positive mixing enthalpy, shown as a function of composition in the top picture of Figure~\ref{enthalpy}.

The input to HydraGNN for each sample includes the three components of the atom position and the proton number. The predicted values include the mixing enthalpy, a single scalar for each sample (graph), as well as the charge transfer and magnitude of the magnetic moment, both scalars per atom (node). Although the magnetic moment is a vector quantity, we treat it as a scalar because all the atomic magnetic moments in the dataset are co-linear 
(all magnetic moments point in the same direction).

The dataset consists of 28,033 configurations out of the $2^{32}$ available, sampled every 3 atomic percent. For this work, if the number of unique configurations for a specific composition is less than 1,000 all those configurations are included in the dataset; for all other compositions, configurations are randomly selected up to 1,000. 
In order to ensure each composition is adequately represented in all portions of the dataset, splitting between the training, validation, and test sets is done separately for each composition.

At the ground state, the total enthalpy $H$ of an alloy is
\begin{equation}
    H = \sum_{i=1}^{E} c_i H_i + \Delta H_{\text{mix}},
\end{equation}
where $E$ is the total number of elements in the system, $c_i$ is the molar fraction of each element $i$, $H_i$ is the molar enthalpy of each element $i$, and $\Delta H_{\text{mix}}$ is the mixing enthalpy. 
We predict the mixing enthalpy for each sample by subtracting the internal enthalpy from the DFT computed total enthalpy as a value more relevant to materials science (more directly related to the configuration).
The chemical disorder makes the task of describing the material properties combinatorially complex; {\textit this represents the main difference from open source databases that have very broad elemental and structural coverage, but only include ordered compounds} \cite{aflow,mp,oqmd}.

The range of values of the mixing enthalpy expressed in Rydberg is $(0.0, 65.92)$, the range of atomic charge transfer in electron charge is $(-5.31, -0.85)$, and the range of atomic magnetic moment in magnetons is $(-0.05, 3.81)$. Since different physical quantities have different units and different orders of magnitude, the inputs and outputs for each quantity are normalized between 0 and 1 across all data.

The total magnetization of the binary alloy FePt is strongly correlated with the concentration of Fe in the alloy, as shown in center picture of Figure \ref{enthalpy}, because only the Fe atoms have a non-negligible magnetic moment. This results in a strong (albeit nonlinear) correlation between the mixing enthalpy and the total magnetization of the alloy, as shown in the bottom picture of Figure \ref{fig:correlation_plots}.

\begin{figure}[h]
    \centering
    \includegraphics[width=0.45\textwidth]{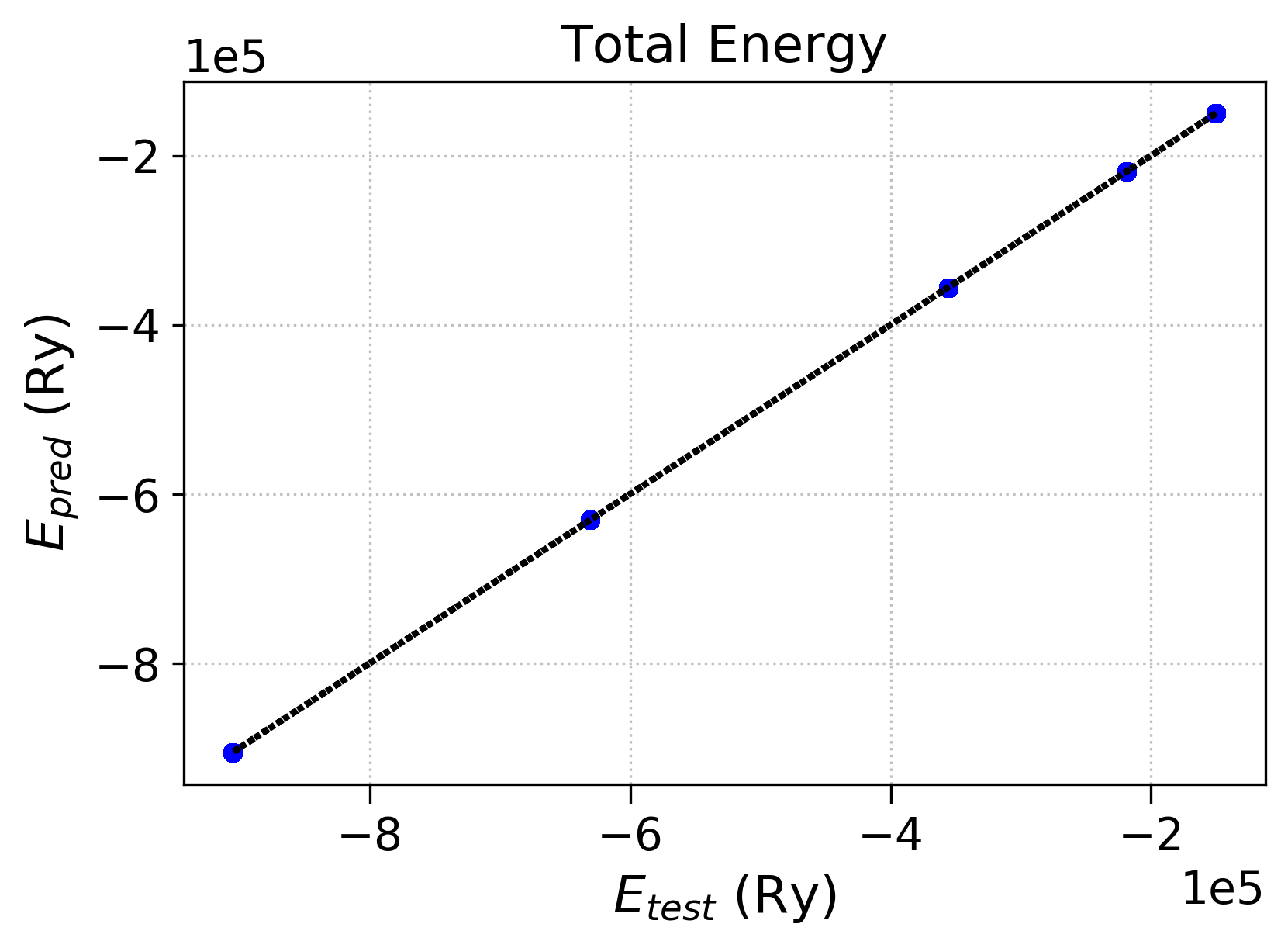}
    \includegraphics[width=0.45\textwidth]{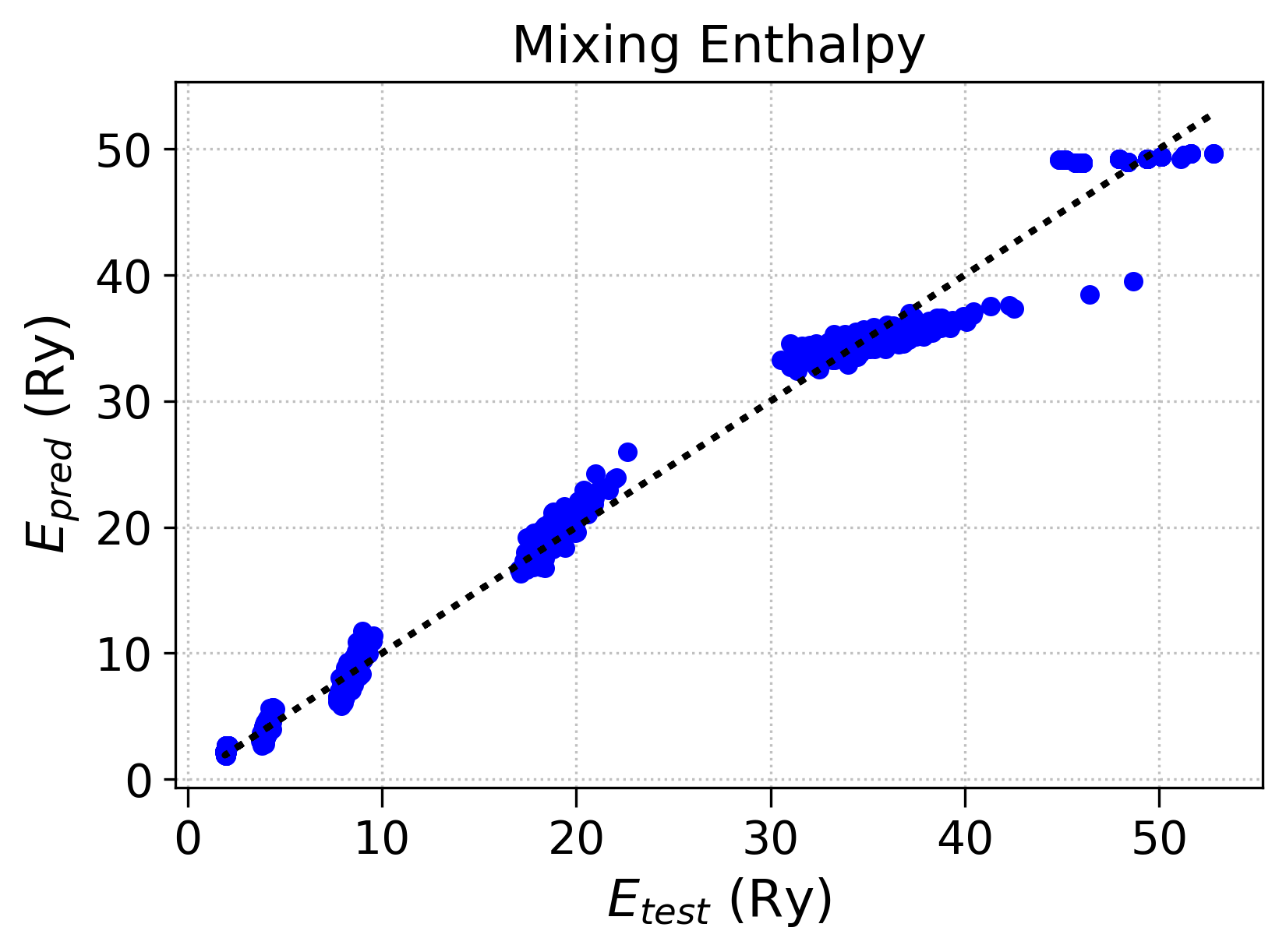}
    \caption{The linear model predictions versus the test set of FePt data for total energy (left) and mixing enthalpy (right). Each cluster of data corresponds to a Fe concentration, ranging from 6.25\% to 93.75\%. }
    \label{fig:linear-model}
\end{figure}

\section{Numerical Section}
This section first presents numerical results using a linear mixing model on portions of the training dataset associated with different compositions of the binary alloys FePt, and the performance of the linear model is interpreted in terms of the physical properties of the ferromagnetic system. 
Additionally, a more complex surrogate model represented by an HydraGNN model is used to simultaneously produce accurate estimates of total mixing enthalpy, atomic charge transfer and atomic magnetic moment over the entire compositional range of the binary alloy. The numerical results that use HydraGNN describe the accuracy of the model as a function of the volume of data used for training.

\subsection{Numerical experiments using linear mixing model}
For binary alloy, there is only one term $V_m^{pp'}$ (see Eq.~\ref{eq:linear}) for each shell, and we consider up to 6th shell. The dataset is randomly splitted with 80\% for training and 20\% for testing, and the model is cross validated among different splits. The reported model performance is evaluated on the test dataset. As shown in Fig.~\ref{fig:linear-model}, the linear model fits perfectly to the total energy, but fails to capture the non-linear behavior of the mixing enthalpy. Since the linear model is designed for the fixed concentration, we first fit the model for each individual Fe concentration (i.e., 6.25\%, 12.5\%, 25\%, 50\%, 75\%, 93.75\%). The observation is that as the Fe concentration increases, the goodness-of-fit deteriorates, indicating the non-linear behavior coming from the magnetism. In Fig.~\ref{fig:linear-model}, we fit the linear model to various Fe concentrations, and each cluster of data corresponds to a Fe concentration, ranging from 6.25\% to 93.75\%.         

\subsection{Numerical experiments using HydraGNN}
\subsubsection*{Training setup}
The architecture of the HydraGNN models has 6 PNA \cite{corso_principal_2020} convolutional layers with 300 neurons per layer. A radius cutoff of 7 $\angstrom$ is used to build the local neighborhoods used by the graph convolutional mask. Every learning task is mapped into separate heads where each head is made up of two fully connected layers, with 50 neurons in the first layer and 25 neurons in the second. Periodic boundary conditions are implemented using the minimum image convention.
The DL models were trained using the Adam method \cite{adam} with a learning rate equal to 0.001, batch sizes of 64, and a maximum number of epochs set to 200. Early stopping is performed to interrupt the training when the validation loss function does not decrease for several consecutive epochs, as this is a symptom that shows further epochs are very unlikely to reduce the value of the loss function.
We reserve 10\% of the total dataset for validation, which corresponds to 2,803 atomic configurations. The remaining 90\% of the total dataset, which amounts to 25,230 atomic configurations, is used to generate different training subsets. 
Each training subset contains a percentage of the total training dataset, ranging from 10\% through 100\% with increments of 10\% across successive data partitions.
 As discussed in Section \ref{dataset_section}, compositional stratified splitting was performed to ensure that all the compositions were equally represented across training, validation, and testing datasets. 
A larger partition contains the smaller ones as subsets. 
The training of the HydraGNN model was performed on one NVIDIA V100 GPU.

\subsubsection*{Performance of HydraGNN on the entire training dataset}
\begin{figure}[h]
    \centering
    \includegraphics[width=0.45\textwidth]{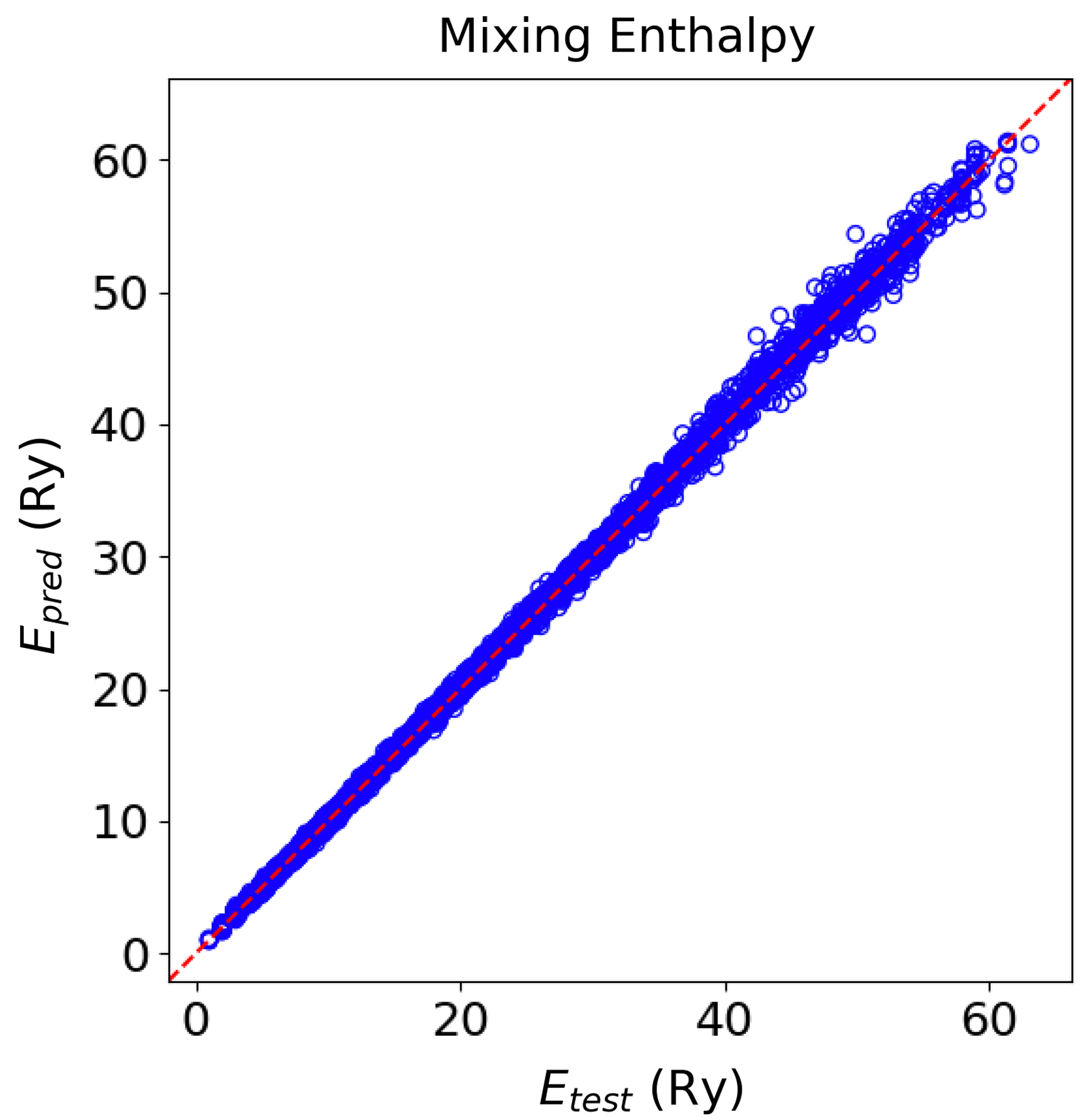}
    \caption{The HydraGNN predictions versus the test set of FePt data for mixing enthalpy. The test set ranges over the entire compositional range by increasing the Fe concentration by 3\%.}
    \label{fig:Hydragnn_model}
\end{figure}

In Figure \ref{fig:Hydragnn_model} we show the parity plot of HydraGNN predictions against the LSMS-3 values for the mixing enthalpy. We notice that HydraGNN accurately predicts the mixing enthalpy for different compositions and atomic configurations associated with each composition, thus clearly outperforming the linear mixing model.  This results corroborate that HydraGNN is an effective non-linear predictive model that accurately reads the dependence of the mixing enthalpy as a function of configurational entropy and chemical composition for a binary ferromagnetic solid solution alloy. 

\subsubsection*{Performance of HydraGNN on different volumes of training data}
\begin{figure}[h]
    \centering
    \includegraphics[width=\textwidth]{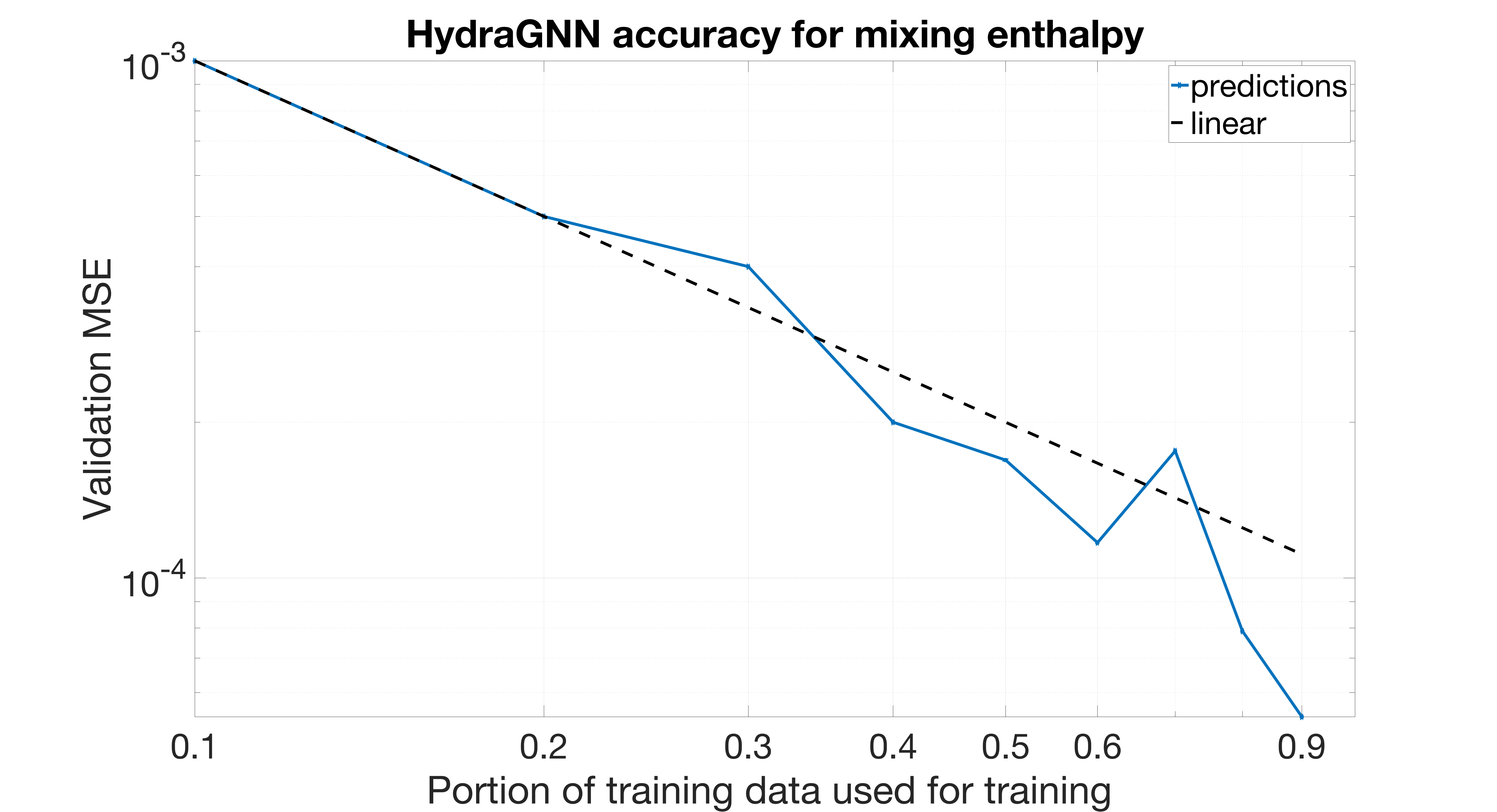}
    \caption{Validation MSE of MTL training performed with HydraGNN for predictions of mixing enthalpy as a function of training data volume used for training.}
    \label{accuracy_hydragnn}
\end{figure}

The validation MSE is used to describe the accuracy of the HydraGNN model as a function of the volume of training data is shown in Figure \ref{accuracy_hydragnn}. The predictive performance shows that the validation MSE decreases linearly with the size of the training data for the mixing enthalpy.

\section{Conclusions and future developments}

In this work we have provided a comparison of two different surrogate models to fit first principles data for alloys. The goal is to provide effective models for the energy of the system that will allow the utilization in Monte-Carlo simulations of the finite temperature statistical mechanics of these systems to enable the quantitative exploration of phase transitions.
The linear model has the significant advantage of faster evaluation and small number of model parameters, which reduces the training effort.
While this linear model had been successfully applied to non-magnetic multi-component alloys \cite{xianglin21} at fixed concentrations, the present work challenges the linear model. As can be seen in Figure \ref{fig:linear-model}, the linear model shows deviation from the correct enthalpy prediction which is dependent on the concentration. Note that this deviation is not immediately apparent in the total energy, which in the all-electron LSMS calculations is dominated by the core-electron binding energy that does not contribute to the material behavior at the energy scale of condensed matter physics.

The use of GCNN allows the description of the interaction for a wide range of concentrations in a single network. As can be seen in Figure \ref{accuracy_hydragnn} the model can be improved over a large range of set sizes in the training data without leading to over-fitting. As the HydraGNN architecture allows us to utilize additional physical information from the LSMS calculations, such as the site magnetization, for a multi-objective optimization of the model using multi-task training, which can lead to better physical prediction of the energy of the system. \cite{pasini}
Thus the HydraGNN model for the FePt system does not suffer the problems of the linear model. In particular, it can capture the changing atomic interactions and their connection to the local Pt magnetic moments at changing atomic concentrations. This greater capability can be attributed  to the inherent non-linear character of NN models that allows more complex functional relationships to be described.
While the advantages come at the cost of greater computational cost when compared to the linear model, the cost is still orders of magnitude lower than the fully selfconsistent DFT calculation that it models.

In summary, we have shown the superior performance of HydraGNN as a surrogate model for predictions of the mixing enthalpy in ferromagnetic solid solution binary alloys. 
%
%

\subsubsection{Acknowledgements} 
This work was supported in part by the Office of Science of the Department of Energy and by the Laboratory Directed Research and Development (LDRD) Program of Oak Ridge National Laboratory. 
This research is sponsored by the Artificial Intelligence Initiative as part of the Laboratory Directed Research and Development (LDRD) Program of Oak Ridge National Laboratory, managed by UT-Battelle, LLC, for the US Department of Energy under contract DE-AC05-00OR22725.
This work used resources of the Oak Ridge Leadership Computing Facility and of the Edge Computing program at the Oak Ridge National Laboratory, which is supported by the Office of Science of the U.S. Department of Energy under Contract No. DE-AC05-00OR22725. 

%
%
%
\bibliographystyle{splncs04}
\bibliography{references}

\begin{thebibliography}{10}
\providecommand{\url}[1]{\texttt{#1}}
\providecommand{\urlprefix}{URL }
\providecommand{\doi}[1]{https://doi.org/#1}

\bibitem{pytorch}
{PyTorch}. \url{https://pytorch.org/docs/stable/index.html}

\bibitem{torch_geometric}
{PyTorch Geometric}. \url{https://pytorch-geometric.readthedocs.io/en/latest/}

\bibitem{CONDUIT2019107644}
Conduit, B., Illston, T., Baker, S., Duggappa, D.V., Harding, S., Stone, H.,
  Conduit, G.: Probabilistic neural network identification of an alloy for
  direct laser deposition. Materials \& Design  \textbf{168},  107644 (2019).
  \doi{https://doi.org/10.1016/j.matdes.2019.107644},
  \url{https://www.sciencedirect.com/science/article/pii/S0264127519300814}

\bibitem{corso_principal_2020}
Corso, G., Cavalleri, L., Beaini, D., Liò, P., Veličković, P.: Principal
  neighbourhood aggregation for graph nets. arXiv:2004.05718 [cs, stat]  (Dec
  2020), \url{http://arxiv.org/abs/2004.05718}, arXiv: 2004.05718

\bibitem{aflow}
Curtarolo, S., Setyawan, W., Hart, G.L.W., Jahnatek, M., Chepulskii, R.V.,
  Taylor, R.H., Wang, S., Xue, J., Yang, K., Levy, O., Mehl, M.J., Stokes,
  H.T., Demchenko, D.O., Morgan, D.: {AFLOW}: {An} automatic framework for
  high-throughput materials discovery. Comput. Mater. Sci.  \textbf{58},
  218--226 (Jun 2012). \doi{10.1016/j.commatsci.2012.02.005},
  \url{http://linkinghub.elsevier.com/retrieve/pii/S0927025612000717}

\bibitem{grain_boundary}
Dai, M., Demirel, M.F., Liang, Y., Hu, J.M.: Graph neural networks for an
  accurate and interpretable prediction of the properties of polycrystalline
  materials. npj computational materials  \textbf{7}(103) (2021).
  \doi{https://doi.org/10.1038/s41524-021-00574-w},
  \url{https://www.nature.com/articles/s41524-021-00574-w#citeas}

\bibitem{GCNNpaper}
Defferrard, M., Bresson, X., Vandergheynst, P.: Convolutional neural networks
  on graphs with fast localized spectral filtering. In: Lee, D., Sugiyama, M.,
  Luxburg, U., Guyon, I., Garnett, R. (eds.) Advances in Neural Information
  Processing Systems. vol.~29. Curran Associates, Inc. (2016),
  \url{https://proceedings.neurips.cc/paper/2016/file/04df4d434d481c5bb723be1b6df1ee65-Paper.pdf}

\bibitem{Bacon1963}
E., B.G., J., C.: Chemical and magnetic order in platinum-rich pt + fe alloys.
  Proc. R. Soc. Lond. A  \textbf{272},  387--405 (1963).
  \doi{10.1098/rspa.1963.0060}

\bibitem{lsms-code}
Eisenbach, M., Li, Y.W., Odbadrakh, O.K., Pei, Z., Stocks, G.M., Yin, J.:
  {LSMS}. https://github.com/mstsuite/lsms,
  \url{https://www.osti.gov//servlets/purl/1420087}

\bibitem{Eisenbach2017}
Eisenbach, M., Larkin, J., Lutjens, J., Rennich, S., Rogers, J.H.: {GPU}
  acceleration of the locally selfconsistent multiple scattering code for first
  principles calculation of the ground state and statistical physics of
  materials. Computer Physics Communications  \textbf{211}, ~2--7 (2017).
  \doi{https://doi.org/10.1016/j.cpc.2016.07.013}

\bibitem{fey_2019}
Fey, M., Lenssen, J.E.: Fast graph representation learning with {PyTorch
  Geometric}. In: ICLR Workshop on Representation Learning on Graphs and
  Manifolds (2019)

\bibitem{10.3389/fmats.2022.865270}
Fuhr, A.S., Sumpter, B.G.: Deep generative models for materials discovery and
  machine learning-accelerated innovation. Frontiers in Materials  \textbf{9}
  (2022). \doi{10.3389/fmats.2022.865270},
  \url{https://www.frontiersin.org/article/10.3389/fmats.2022.865270}

\bibitem{hamilton2017inductive}
Hamilton, W.L., Ying, R., Leskovec, J.: Inductive representation learning on
  large graphs. In: Proceedings of the 31st International Conference on Neural
  Information Processing Systems. pp. 1025--1035 (2017)

\bibitem{Higashiyama2003}
Higashiyama, Y., Tsunoda, Y.: Magnetism of pt0.67fe0.33 alloy. Journal of the
  Physical Society of Japan  \textbf{72}(12),  3305--3306 (2003).
  \doi{10.1143/JPSJ.72.3305}

\bibitem{ja_bondy_usr_murty_graphs_nodate}
J.A.~Bondy, U.M.: Graphs and subgraphs. In: Graph theory with applications.
  North-Holland

\bibitem{mp}
Jain, A., Ong, S.P., Hautier, G., Chen, W., Richards, W.D., Dacek, S., Cholia,
  S., Gunter, D., Skinner, D., Ceder, G., Persson, K.A.: Commentary: {The}
  materials project: {A} materials genome approach to accelerating materials
  innovation. APL Mater.  \textbf{1}(1),  0--11 (2013). \doi{10.1063/1.4812323}

\bibitem{adam}
Kingma, D.P., Ba, J.: Adam: a method for stochastic optimization.
  arXiv:1412.6980 [cs]  (Jan 2017), \url{http://arxiv.org/abs/1412.6980},
  arXiv: 1412.6980

\bibitem{Lavrentiev}
Lavrentiev, M.Y., Drautz, R., Nguyen-Manh, D., Klaver, T.P.C., Dudarev, S.:
  Monte carlo study of thermodynamic properties and clustering in the {BCC}
  {Fe-Cr} system. Physical Review B  \textbf{75}(014208) (2007).
  \doi{10.1103/PhysRevB.75.014208}

\bibitem{xianglin21}
Liu, X., Zhang, J., Yin, J., Bi, S., Eisenbach, M., Wang, Y.: Monte carlo
  simulation of order-disorder transition in refractory high entropy alloys: A
  data-driven approach. Computational Materials Science  \textbf{187},  110135
  (2021). \doi{https://doi.org/10.1016/j.commatsci.2020.110135}

\bibitem{gatgnn}
Louis, S.Y., Zhao, Y., Nasiri, A., Wang, X., Song, Y., Liu, F., Hu, J.: Graph
  convolutional neural networks with global attention for improved materials
  property prediction. Phys. Chem. Chem. Phys.  \textbf{22},  18141--18148
  (2020). \doi{10.1039/D0CP01474E}, \url{http://dx.doi.org/10.1039/D0CP01474E}

\bibitem{lupo_gcnn}
Lupo~Pasini, M., Bur\^cul, M., Reeve, S.T., Eisenbach, M., Perotto, S.: Fast
  and accurate predictions of total energy for solid solution alloys with graph
  convolutional neural networks. Springer Journal of Communications in Computer
  and Information Science  \textbf{1512} (Sep 2021)

\bibitem{FePt}
Lupo~Pasini, M., Eisenbach, M.: {FePt binary alloy with 32 atoms - LSMS-3 data
  - DOI:10.13139/OLCF/1762742}  (2 2021). \doi{10.13139/OLCF/1762742},
  \url{https://www.osti.gov/dataexplorer/biblio/dataset/1762742}

\bibitem{pasini}
Lupo~Pasini, M., Li, Y.W., Yin, J., Zhang, J., Barros, K., Eisenbach, M.: Fast
  and stable deep-learning predictions of material properties for solid
  solution alloys. J. Phys.: Condens. Matter  \textbf{33}(8),  084005 (Dec
  2020). \doi{10.1088/1361-648X/abcb10},
  \url{https://doi.org/10.1088/1361-648x/abcb10}, publisher: IOP Publishing

\bibitem{Lupo_Pasini_2022}
Lupo~Pasini, M., Zhang, P., Reeve, S.T., Choi, J.Y.: Multi-task graph neural
  networks for simultaneous prediction of global and atomic properties in
  ferromagnetic systems. Machine Learning: Science and Technology
  \textbf{3}(2),  025007 (may 2022). \doi{10.1088/2632-2153/ac6a51},
  \url{https://doi.org/10.1088/2632-2153/ac6a51}

\bibitem{hydragnn}
Lupo~Pasini, M., Reeve, S.T., Zhang, P., Choi, J.Y.: {HydraGNN}. [Computer
  Software] \url{https://doi.org/10.11578/dc.20211019.2} (oct 2021).
  \doi{10.11578/dc.20211019.2}, \url{https://github.com/ORNL/HydraGNN}

\bibitem{doi:10.1080/14786435.2013.805275}
Lépinoux, J., Sigli, C.: Precipitate growth in concentrated binary alloys: a
  comparison between kinetic monte carlo simulations, cluster dynamics and the
  classical theory. Philosophical Magazine  \textbf{93}(23),  3194--3215
  (2013). \doi{10.1080/14786435.2013.805275},
  \url{https://doi.org/10.1080/14786435.2013.805275}

\bibitem{MOHAMMADI20209620}
Mohammadi, H., Eivani, A.R., Seyedein, S.H., Ghosh, M.: Modified monte carlo
  approach for simulation of grain growth and ostwald ripening in two-phase
  zn–22al alloy. Journal of Materials Research and Technology  \textbf{9}(5),
   9620--9631 (2020). \doi{https://doi.org/10.1016/j.jmrt.2020.06.017},
  \url{https://www.sciencedirect.com/science/article/pii/S2238785420314186}

\bibitem{pytorch2019}
Paszke, A., Gross, S., Massa, F., Lerer, A., Bradbury, J., Chanan, G., Killeen,
  T., Lin, Z., Gimelshein, N., Antiga, L., Desmaison, A., Kopf, A., Yang, E.,
  DeVito, Z., Raison, M., Tejani, A., Chilamkurthy, S., Steiner, B., Fang, L.,
  Bai, J., Chintala, S.: Pytorch: An imperative style, high-performance deep
  learning library. In: Wallach, H., Larochelle, H., Beygelzimer, A.,
  d\textquotesingle Alch\'{e}-Buc, F., Fox, E., Garnett, R. (eds.) Advances in
  Neural Information Processing Systems 32, pp. 8024--8035. Curran Associates,
  Inc. (2019),
  \url{http://papers.neurips.cc/paper/9015-pytorch-an-imperative-style-high-performance-deep-learning-library.pdf}

\bibitem{Reitz}
Reitz, D.M., Blaisten-Barojas, E.: Simulating the nak eutectic alloy with monte
  carlo and machine learning. Scientific Reports  \textbf{9}(704) (2019).
  \doi{doi.org/10.1038/s41598-018-36574-y}

\bibitem{oqmd}
Saal, J.E., Kirklin, S., Aykol, M., Meredig, B., Wolverton, C.: Materials
  design and discovery with high-throughput density functional theory: the
  {Open} {Quantum} {Materials} {Database} ({OQMD}). JOM  \textbf{65}(11),
  1501--1509 (Nov 2013). \doi{10.1007/s11837-013-0755-4},
  \url{http://link.springer.com/10.1007/s11837-013-0755-4}

\bibitem{GNNpaper}
Scarselli, F., Gori, M., Tsoi, A.C., Hagenbuchner, M., Monfardini, G.: The
  graph neural network model. IEEE Transactions on Neural Networks
  \textbf{20}(1),  61--80 (2009). \doi{10.1109/TNN.2008.2005605}

\bibitem{Slater1936}
Slater, J.C.: The ferromagnetism of nickel. Phys. Rev.  \textbf{49},  537--545
  (Apr 1936). \doi{10.1103/PhysRev.49.537}

\bibitem{Tetot1996}
T{\'e}tot, R., Finel, A.: Relaxed Monte Carlo Simulations on Au-Ni Alloy, pp.
  179--184. Springer US, Boston, MA (1996). \doi{10.1007/978-1-4613-0385-5_8},
  \url{https://doi.org/10.1007/978-1-4613-0385-5_8}

\bibitem{Tobita2010}
Tobita, N., Nakajima, N., Ishimatsu, N., Maruyama, H., Shimada, K., Namatame,
  H., Taniguchi, M.: Antiferromagnetic phase transition in ordered fept3
  investigated by angle-resolved photoemission spectroscopy. Journal of the
  Physical Society of Japan  \textbf{79}(2),  024703 (2010).
  \doi{10.1143/JPSJ.79.024703}

\bibitem{Vlaic2010}
Vlaic, P., Burzo, E.: Magnetic behaviour of iron-platinum alloys. Journal of
  Optoelectronics and Advanced Materials  \textbf{12},  1114--1124 (05 2010)

\bibitem{Walle2002SelfdrivenLM}
van~de Walle, A., Asta, M.: Self-driven lattice-model monte carlo simulations
  of alloy thermodynamic properties and phase diagrams. Modelling and
  Simulation in Materials Science and Engineering  \textbf{10},  521--538
  (2002)

\bibitem{Wang1995}
Wang, Y., Stocks, G.M., Shelton, W.A., Nicholson, D.M.C., Temmerman, W.M.,
  Szotek, Z.: Order-{N} multiple scattering approach to electronic structure
  calculations. Phys. Rev. Lett.  \textbf{75}, ~2867 (1995).
  \doi{https://doi.org/10.1103/PhysRevLett.75.2867}

\bibitem{cgcnn}
Xie, T., Grossman, J.C.: Crystal graph convolutional neural networks for an
  accurate and interpretable prediction of material properties. Phys. Rev.
  Lett.  \textbf{120}(14),  145301 (Apr 2018).
  \doi{10.1103/PhysRevLett.120.145301},
  \url{https://link.aps.org/doi/10.1103/PhysRevLett.120.145301}

\end{thebibliography}

\end{document}